\documentclass[preprint,3p,twocolumn]{elsarticle}

\usepackage{amsmath,graphicx,color}
\usepackage{booktabs,array,dcolumn}

\usepackage{multirow}
\usepackage{times}
\newcolumntype{C}{>{\centering}p}

\begin{document}

\title{On the performance of interatomic potential models of iron: comparison of the phase diagrams}

\author[lbp]{L\'\i via B. P\'artay}
\address[lbp]{Department of Chemistry, University of Reading, Reading RG6 6AD, UK}
\ead{l.bartokpartay@reading.ac.uk}

\date{\today}

\begin{abstract}
In order to study the performance of interatomic potentials and their reliability at higher pressures,  the phase diagram of four different embedded-atom type potential models of iron is compared. 
The calculations were done by the nested sampling technique in the pressure range 0.1~GPa--100~GPa. 
The low pressure stable structure is found to be the body-centred cubic in all cases, but the higher pressure phases show a great variation, being face-centred cubic, hexagonal close-packed and -- at very low temperatures -- different body-centred tetragonal phases are observed as well.
The melting line is overestimated considerably for three of the models, but for the one where liquid properties had been taken into account during the potential fitting process, the agreement with experimental results is good, even at very high pressures. 
\end{abstract}


\maketitle


\section{Introduction}

When using atomistic simulations, one of the keys to accurately predict the structure and properties of materials and their defects is the quality of the description of atomic interactions. 
In order to compromise between computational efficiency, generality and accuracy, empirical or semi-empirical potentials are often used as descriptors in most large-scale and long-time computations.
These potentials are commonly determined by fitting a proposed functional form to a group of available data, which may be obtained from either experimental measurements or first-principles calculations. 

However, it is difficult to predict how these potentials will perform under conditions different from that of the exact fitting parameters.
While some of the microscopic properties are likely to be reproduced accurately, the macroscopic behaviour of the potential model can be very different from what is expected, especially the complex phenomena of phase transitions and phase stability.
Gaining a better understanding on how the choice of fitting parameters effect the resulting phase diagram can help us not only to determine the reliability of a certain potential model, but to improve strategies of potential development in the future.  
To be able to do this, it is vital to have a technique which is capable of calculating the entire phase diagram of a potential model in an automated way, without prior knowledge of the phases. 
In recent years we have been developing such a technique, called nested sampling (NS).
NS is a Bayesian statistical method~\cite{bib:skilling,bib:skilling2} we adapted to explore the atomic phase space~\cite{our_NS_paper} and has been applied to study clusters and the hard-sphere model ~\cite{our_NSHS_paper, diffns,Frenkel_NS}.
In previous papers~\cite{pt_phase_dias_ns,pymatnest_paper} we also showed how the NS algorithm enables the automated calculation of the complete pressure-temperature-composition phase diagram.
In the present work I further demonstrate this, using empirical potentials of iron and discussing the emerging differences between the studied models.

\subsection{Experimental phase diagram of iron} 

Iron is one of the most important and widely used technological materials, moreover, it is considered to be the dominant component of the inner core of the Earth. 
Thus, investigating the properties and phase behaviour of iron is not only of great industrial importance but fundamental in the understanding of geological processes and the inner structure of our planet.
However, due to its unique properties, the complex phase diagram of iron is still not fully understood, with details of the melting line and crystal structures at high pressure being in the focus of research for several decades.


At low temperature and pressure iron exists in the $\alpha$-Fe form, a ferromagnetic bcc structure. 
With increasing temperature this first transforms to the non-magnetic fcc structure, $\gamma$-Fe, then to a ferromagnetic bcc crystal again, called $\delta$-Fe~\cite{Strong_delta_exp}.
At higher pressures the hcp structure ($\epsilon$-Fe) becomes stable,
the triple point among the bcc, hcp, and fcc structures located at about 11~GPa and 750~K~\cite{Fe_structure}.
The ground state structure is predicted to remain hcp up to about 300~GPa, however the phase diagram and crystal structure properties are well defined only up to 20~GPa.
While the transition between the $\gamma$-Fe (fcc) and $\epsilon$-Fe (hcp) at temperatures close to the melting is usually seen at approx. 50~GPa~\cite{Yoo_solid_exp}, some measurements indicate that this happens at significantly higher pressures~\cite{ge_triplepoint}.
First principle calculations predict that at high temperature and above 300~GPa, the stability of the fcc phase becomes comparable again to that of the hcp, suggesting that at extreme pressures both phases might be stable as well as other close packed stacking sequences~\cite{dhcp_Fe2}.
Furthermore, it has been suggested that two more stable crystal structures might exist:
there is evidence about the existence of a double-hcp structure, called $\epsilon'$-Fe between 15~GPa-40~GPa~\cite{dhcp_Fe1}, and speculations about a high temperature $\beta$-Fe phase (of unknown structure) at pressures above 50~GPa~\cite{Saxena_melt_exp,ge_triplepoint,Saxena_melt2_exp}.

The agreement between the melting temperatures measured by different experimental techniques is good up to 20~GPa, though above that there is a discrepancy between the results, with the suggested melting temperature ranging from 2800~K to 4100~K at 100~GPa~\cite{Williams_melt_exp,static_melt_exp,static_melt_exp2,static_melt_exp3,Saxena_melt_exp,ge_triplepoint,melt_Fe}.

\subsection{Potential models for iron}

In order to gain a better understanding on the properties of iron, different computer simulation and modelling techniques are regularly used to study e.g. the crystal stability, surface properties, defects or radiation damage. 
Several interatomic potential models have been developed in the past decades to allow large scale calculations, especially within the embedded atom model (EAM) framework~\cite{ironpot_osetsky, ironpot_Lee, Ackland_potential,MM_potential,Chamati_potential,Malerba_potential1}. 
 
One of the most widely used EAM potentials for iron was developed by Ackland \textit{et al}.~\cite{Ackland_potential}. 
The coefficients of this potential were chosen to fit the lattice parameter, cohesive energy, unrelaxed vacancy formation energy and elastic constants for  $\alpha$-Fe at $T=0$K. I will refer to this potential as \emph{Ackland97}. 
This potential is known to overestimate the melting point by at least 500~K at ambient pressure and provides a liquid structure that is more ordered than observed experimentally~\cite{MM_potential}. 

A set of potentials were developed by Mendelev \textit{et al.}~\cite{MM_potential} with liquid parameters taken into account during the fitting procedure. 
As a result these EAM models reproduce well the melting data and liquid structure factor of iron.
I chose the recommended potential Nr.2 to use in the current study and will refer to this potential as \emph{Mendelev03}.

Chamati \textit{et al.} published a model~\cite{Chamati_potential}, where both DFT and experimental data were included in the parameter fit (elastic constant, vacancy formation), including the energies of bcc, fcc, simple cubic and diamond cubic structures. 
The potential accurately predicts bulk and surface properties for both bcc and fcc iron, as well as describing the surface migration, phonon dispersion curves and thermal expansion properties, though these were not included while fitting the potential parameters.
I will refer to this potential as \emph{Chamati06}.

Another EAM potential was developed by Marinica \textit{et al.}, where although liquid properties were not included in the fit, DFT defect formation and migration energies were~\cite{Malerba_potential1,Malerba_potential2}. 
I will refer to this potential as \emph{Marinica07}.

It has to be noted that these EAM models lack the description of magnetic properties which play a significant role in solid-solid transitions\cite{Bagno_Fe,magn_hcp}, moreover, due to the general formalism of the EAM potential, the interactions are spherically symmetric, missing the effect of directionality caused by the partially filled $d$ bands of iron. 
More complex interaction models are capable of providing a more accurate description of phase behaviour, such as MEAM~\cite{ironpot_Lee}, bond-order potentials which can predict the bcc-hcp phase transition correctly~\cite{ironpot_genrich} and the transition sequence of $\alpha - \beta - \gamma$ phases~\cite{ironpot_muller}, or magnetic bond-order potentials, which reproduces accurately the relative stability of different magnetic bulk phases~\cite{ironpot_mrovec}. 
Nevertheless, due to their relative simplicity and low computational cost, EAM potentials are extensively used for studying different properties of iron, thus understanding the phase behaviour of these models is an important information
for a wide range of applications.


\section{Computational details}

The nested sampling calculations were performed as presented in~\cite{pt_phase_dias_ns}. 
The simulations were run at constant pressure, and the simulation cell of variable shape and size contained 64 atoms. 
The initial configurations were generated randomly. 
New samples were generated with performing Hamiltonian Monte Carlo~\cite{pymatnest_paper} (all-atom) moves, and changing the volume and the shape of the cell by shear and stretch moves. 
Overall 800 steps were performed at every iteration with ratio 1:2:2:2, respectively.
Parallel implementations of this algorithm is available in the {\tt pymatnest}  python software package~\cite{pymatnest}, using the
LAMMPS package~\cite{LAMMPS} for the dynamics.

The number of walkers were chosen such that the difference in the melting temperature predicted by independent parallel runs is less than 100~K, and the solid-solid transition is reliably found. 
At pressures where there are no solid-solid transition, 640 or 1280 walkers were enough to reach convergence, and when solid-solid transition were present 1920 walkers were needed. 
The exception is the high pressure hcp-fcc transition of the Chamati06 potential, where 4800 walkers were needed to consistently see the transition. 
The computational cost associated with different number of walkers is shown in Table~\ref{table:comp_param}.
The error bars reported on the phase diagrams correspond to the full width at half maximum of the heat capacity peaks.  

When the sampling process was finished, the heat capacity curves were calculated, its peak positions allowing us to draw the phase diagram. 
To aid the identification of the solid structure and determine the solid-solid phase transitions both the bond order parameters~\cite{bib:Q6parameter} and the weighted average of radial distribution functions were used.

\begin{table}[htb]
\begin{center}
\caption{Typical number of walkers and the associated computational cost. Total force evaluations are counted as for 64 atoms.}
\begin{tabular}{ccc}
\hline\hline
  number  &  total force  & CPU (h)  \\  %
 of walkers & evaluations &                  \\ 
\hline
 640   & $9.6\times10^8$  & 640  \\
 1280 & $1.9\times10^9$  & 1280  \\
 1920 & $3.8\times10^9$  & 3460  \\
 4800 & $8.6\times10^9$  & 8820  \\
\hline\hline
\label{table:comp_param}
\end{tabular}
\end{center}
\end{table}

\section{Results}

\subsection{potential Ackland97}

\begin{figure}[hbt]
\begin{center}
\includegraphics[angle=90,width=8cm]{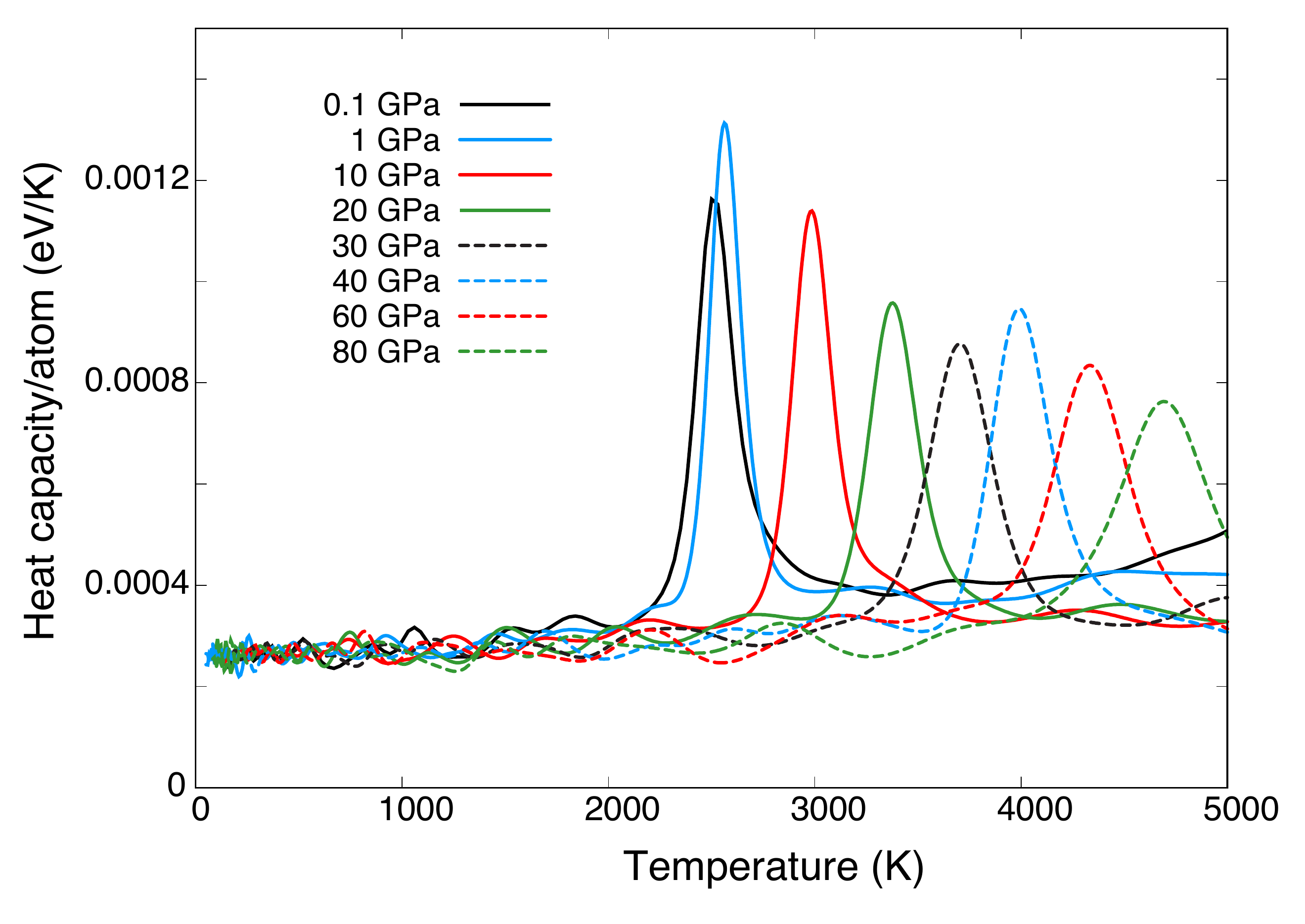}
\end{center}
\vspace{-20pt}
\caption {Heat capacity curves of the Ackland97 potential at different pressures. The peaks correspond to the melting transition.}
\label{fig:Ackaland_cp}
\end{figure}

\begin{figure}[h!]
\begin{center}
\includegraphics[width=7cm]{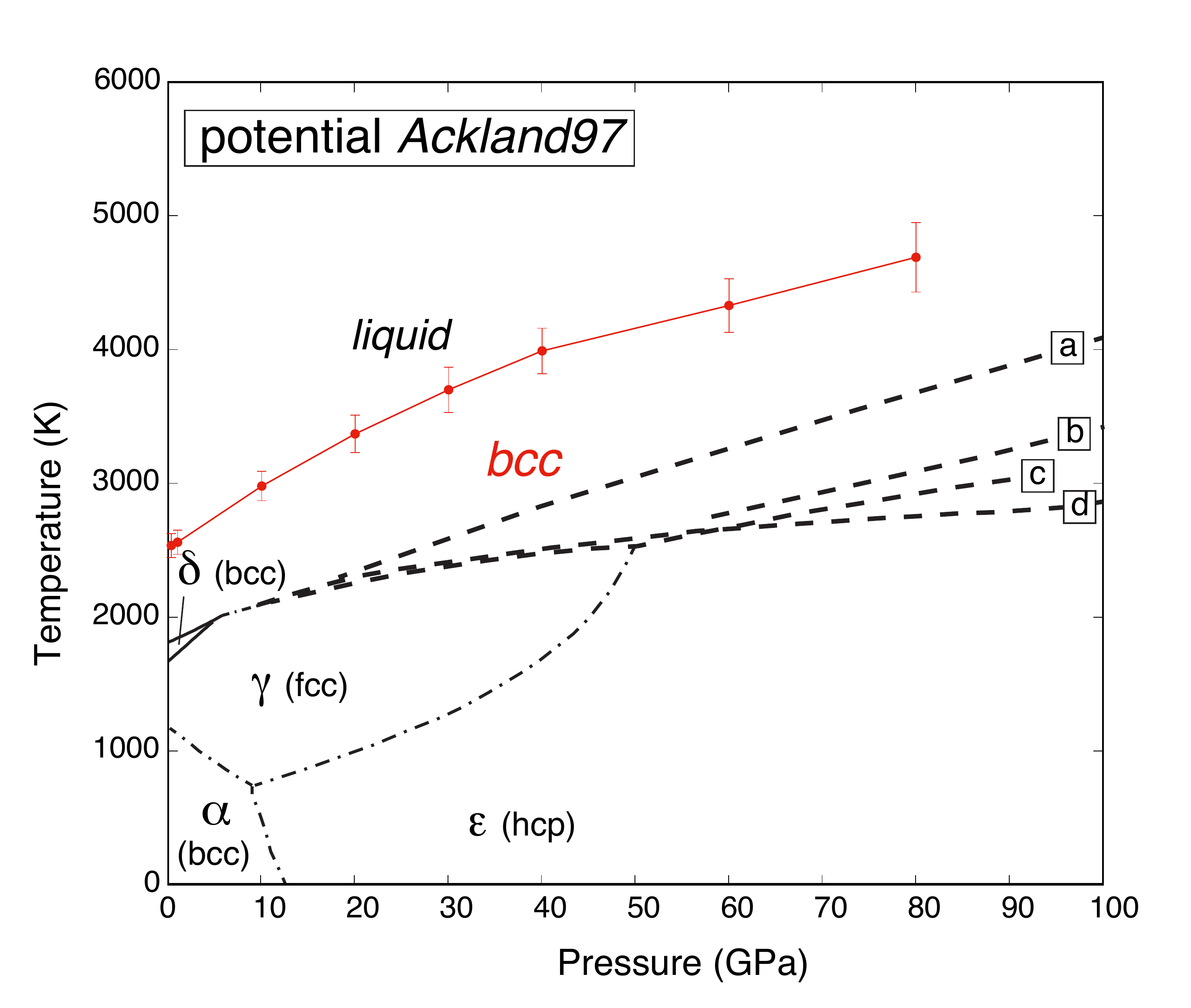}
\end{center}
\vspace{-20pt}
\caption {Phase diagram of iron with the Ackland97 potential. Black lines correspond to experimental data (long dash lines correspond to melting data: (a)\cite{Williams_melt_exp},  (b)\cite{static_melt_exp,static_melt_exp2,static_melt_exp3}, (c)\cite{Saxena_melt_exp} and (d)\cite{ge_triplepoint},while solid-solid transitions are shown by dash-dot lines~\cite{Yoo_solid_exp} and solid lines~\cite{Strong_delta_exp}. 
Suspected $\epsilon'$ and $\beta$ phases are not shown). Red symbols and lines correspond to the Ackland97 potential results.}
\label{fig:Ackland_pd}
\end{figure}

Heat capacity curves and the phase diagram of the Ackland97 potential are shown in Figure~\ref{fig:Ackaland_cp} and \ref{fig:Ackland_pd}, respectively. 
The melting line is overestimated considerably, with the low pressure melting point being about 700~K higher than the experimental value, with the same trend continuing at higher pressures.
The only stable solid structure is bcc in the studied pressure range, but the fcc phase, which is metastable for this model, has also been studied~\cite{iron_melt_interf}. 

\subsection{potential Mendelev03}

\begin{figure}[hbt]
\begin{center}
\includegraphics[width=7cm]{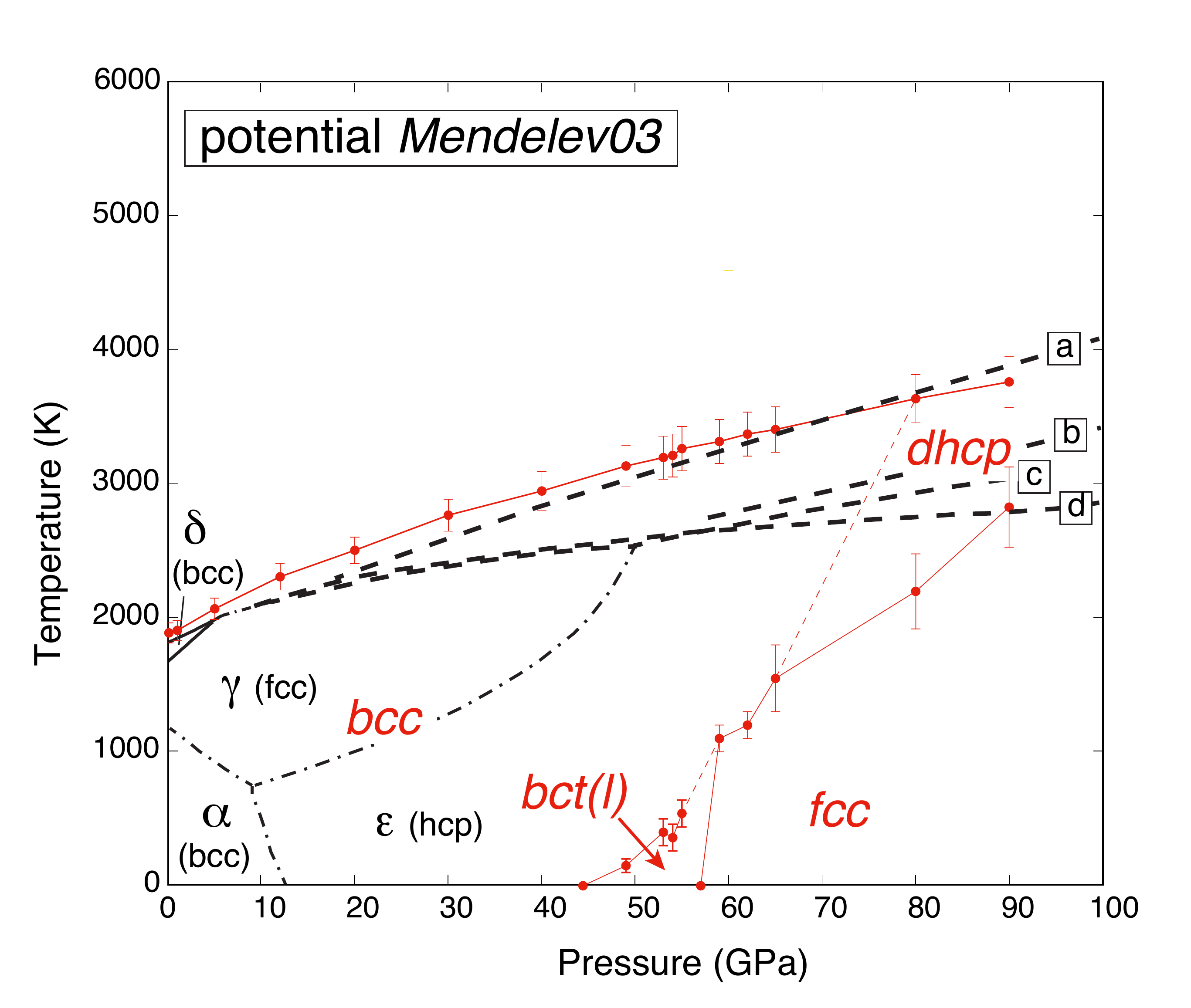}
\end{center}
\vspace{-20pt}
\caption {Phase diagram of iron with the Mendelev03 potential. Black lines correspond to experimental data (long dash lines correspond to melting data: (a)\cite{Williams_melt_exp},  (b)\cite{static_melt_exp,static_melt_exp2,static_melt_exp3}, (c)\cite{Saxena_melt_exp} and (d)\cite{ge_triplepoint},while solid-solid transitions are shown by dash-dot lines~\cite{Yoo_solid_exp} and solid lines~\cite{Strong_delta_exp}. Suspected $\epsilon'$ and $\beta$ phases are not shown). Red symbols and lines correspond to the Mendelev03 potential results.}
\label{fig:MM_pd}
\end{figure}

The calculated phase diagram of the Mendelev03 potential can be seen in Figure~\ref{fig:MM_pd}.

As shown before~\cite{MM_potential}, the melting temperature agrees very well with the experimental value, 
but more importantly this agreement is still excellent at much higher pressures, and fits the experimental data of Williams et al.\cite{Williams_melt_exp} up to 90~GPa. 
This suggests that including low pressure liquid structure properties in the potential fit can improve the reliability of the predicted melting line in a wide range, even up to very high pressures.

\begin{figure}[hbt]
\begin{center}
\includegraphics[angle=90,width=7cm]{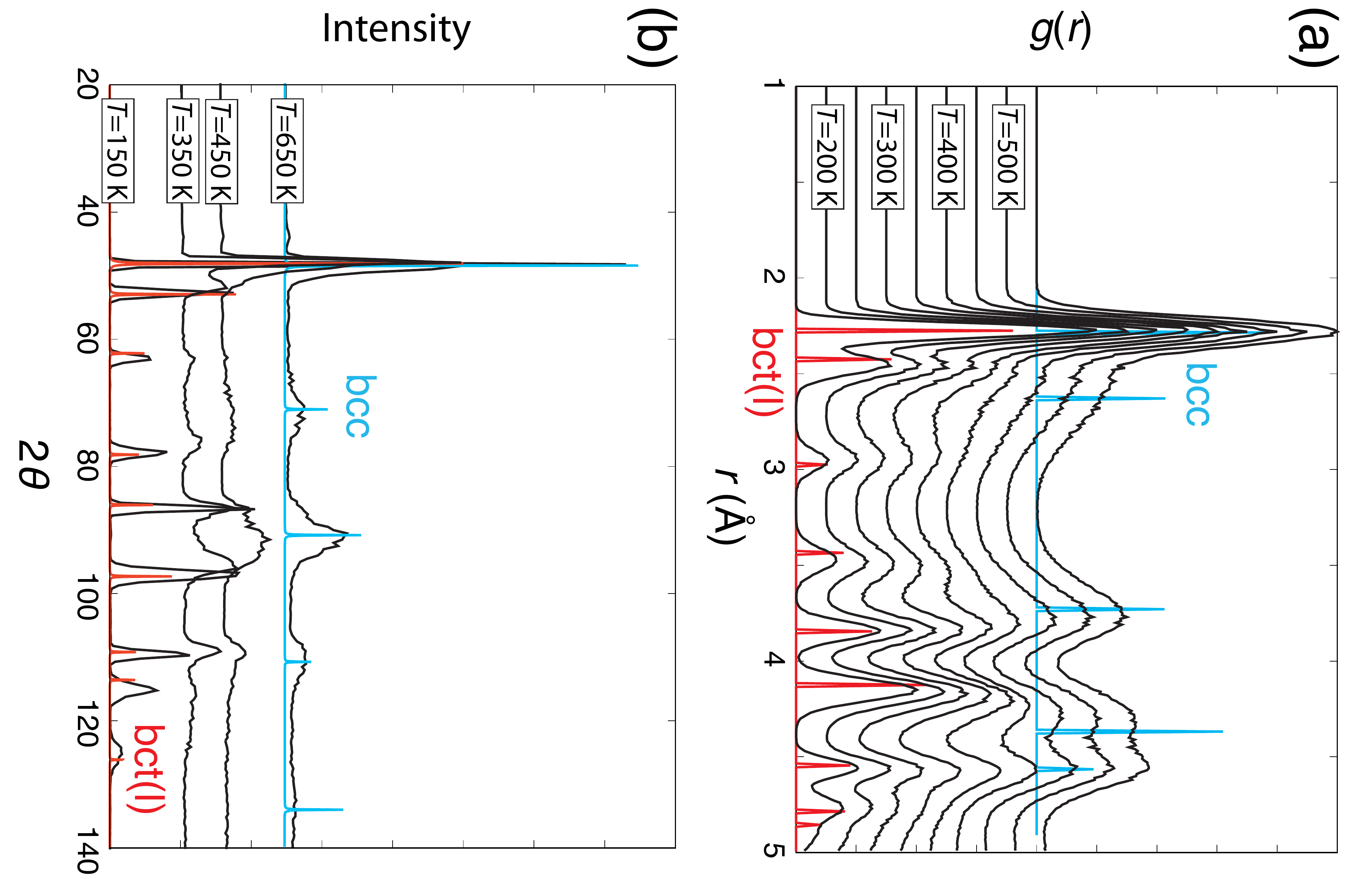}
\end{center}
\vspace{-20pt}
\caption {Mendelev03 potential at $p=54$~GPa. (a) Weighted average radial distribution functions, and (b) weighted average powder diffraction spectra at different temperatures. The red and blue lines show the reference results of the bct(l) and bcc structures, respectively. The curves are shifted vertically for clarity.}
\label{fig:MM_gr}
\end{figure}

At lower pressures the stable structure is bcc, then between 44~GPa and 57~GPa the body centred tetragonal structure, bct, becomes the ground state structure. 
Compared to the bcc, the cell is elongated to one direction ($c/a=1.22$) and this allows the packing fraction, thus the density to be increased by~2.8\%. 
This explains why this structure becomes favourable at higher pressures. 
Figure~\ref{fig:MM_gr} shows the weighted average radial distribution functions and powder diffraction spectra at different temperatures at 54~GPa. 
The additional small peaks corresponding to the bct(l) structure disappears between 350~K and 400~K, indicating that above that the bcc structure is more favourable than the slightly elongated bct(l).

\begin{figure}[hbt]
\begin{center}
\includegraphics[width=5cm]{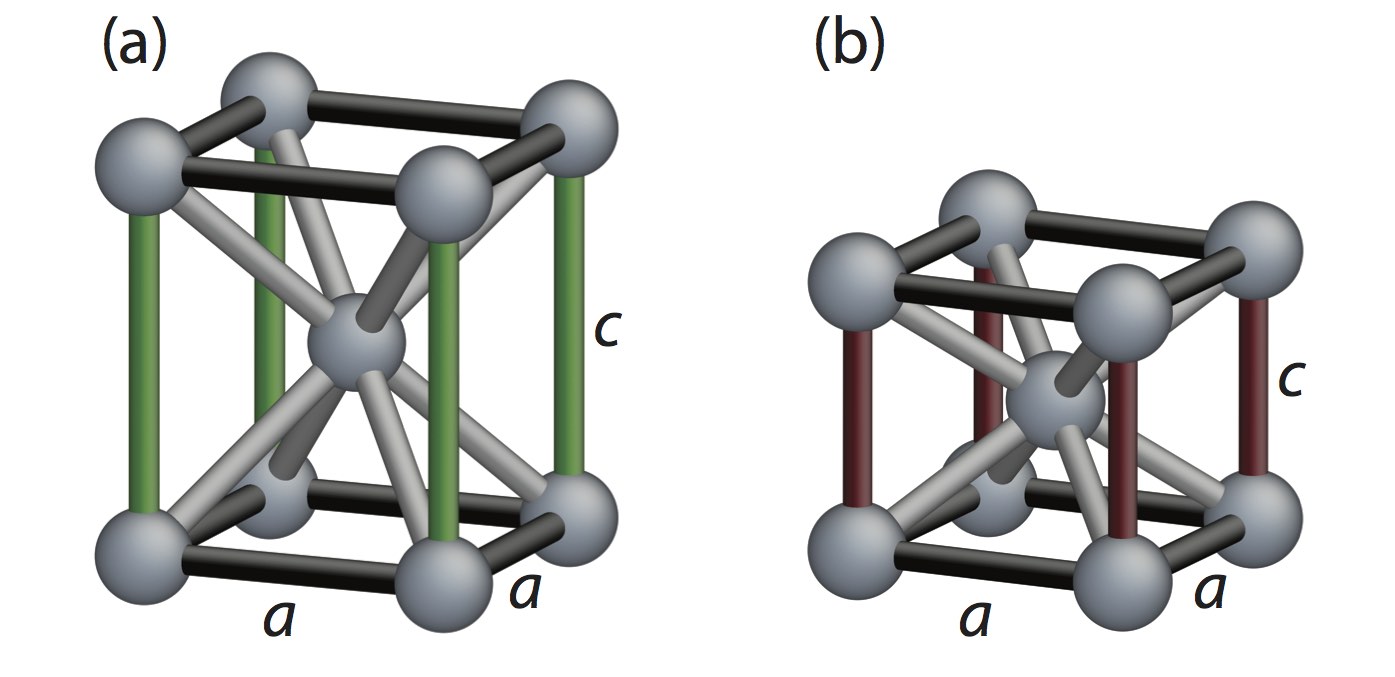}
\end{center}
\vspace{-20pt}
\caption {Structures of the body-centred-tetragonal structure seen in simulations. (a) elongated tetragonal cell $c/a>1.0$ referred to as bct(l) and (b) shortened tetragonal cell, $c/a<1.0$, referred to as bct(s).}
\label{fig:bct_snap}
\end{figure}
 
\begin{figure}[hbt]
\begin{center}
\includegraphics[angle=90,width=7cm]{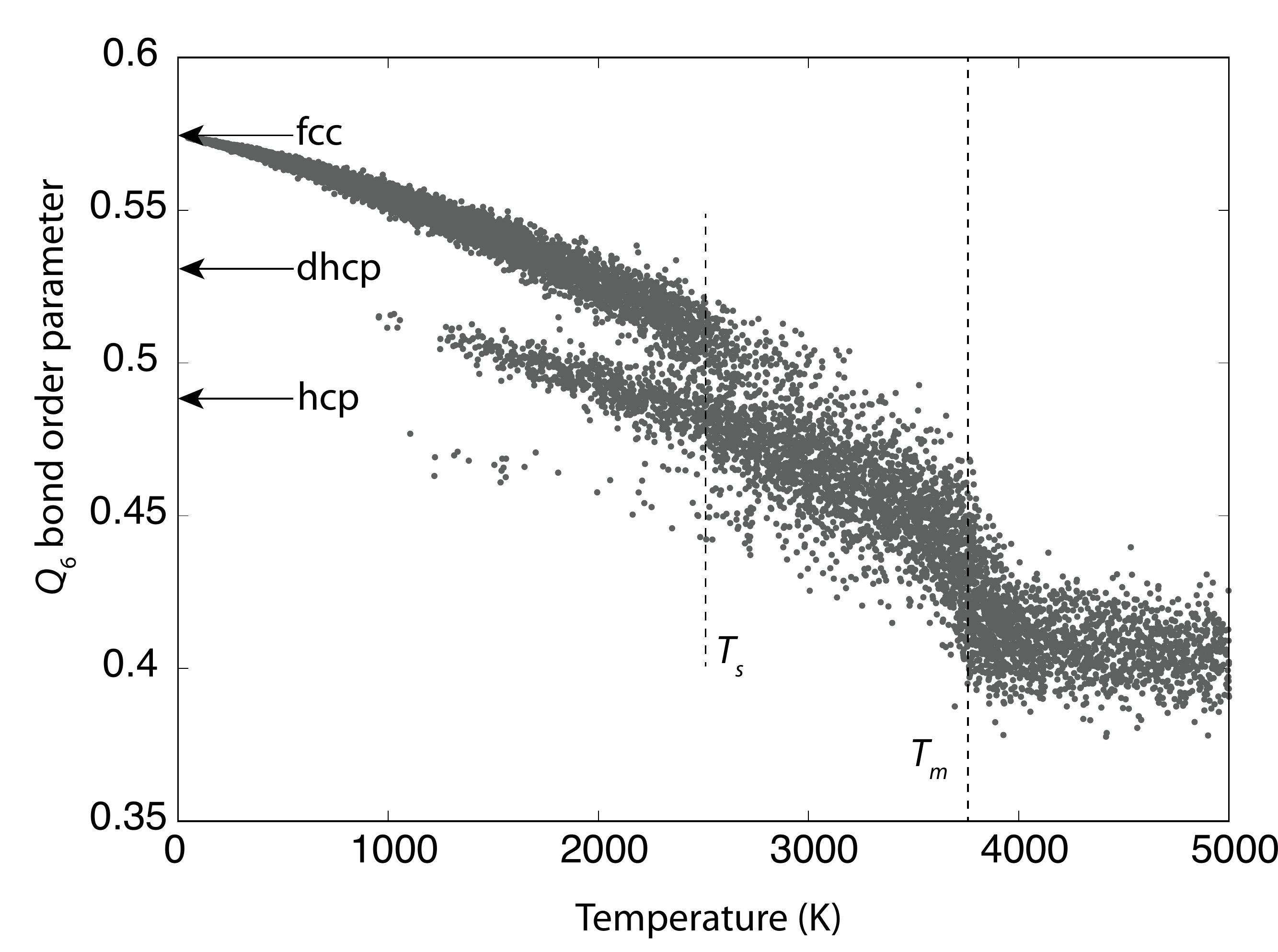}
\end{center}
\vspace{-20pt}
\caption {$Q_6$ bond order parameter of the configurations as a function of temperature, generated during a nested sampling run with the Mendelev03 potential at 90~GPa. The dashed lines correspond to phase transitions and the $Q_6$ of some perfect crystal structures are marked by arrows.}
\label{fig:MM90_qw}
\end{figure}

As the pressure is increased above 57~GPa, the fcc structure becomes the ground state (which can be thought of as the same as a bct structure with $c/a=\sqrt{2}$, thus a bct with an even higher density).
The high temperature phase also changes from bcc to a mixed stacking close packed structure, containing hexagonal (h) and cubic (c) layers in the same ratio, such as $\langle \mathrm{hc} \rangle$ (known as double-hexagonal close packed, dhcp) and $\langle \mathrm{hhcc}\rangle$. 
As the pressure increases further, more and more configurations of dhcp structure are seen, and at 90GPa, hcp configurations begin to be formed too, although this accounts for the minority of the structures observed. 
This is demonstrated on Figure~\ref{fig:MM90_qw}, where the average $Q_6$ bond order parameter of the configurations generated during a NS calculation at 90~GPa are shown. Below the melting temperature the majority of the configurations have an order parameter corresponding to the dhcp structure, but fcc and hcp are present as well. At 2500~K the ratio of fcc structures becomes larger indicating that it becomes the favourable solid state.


\subsection{potential Marinica07}

\begin{figure}[hbt]
\begin{center}
\includegraphics[width=7cm]{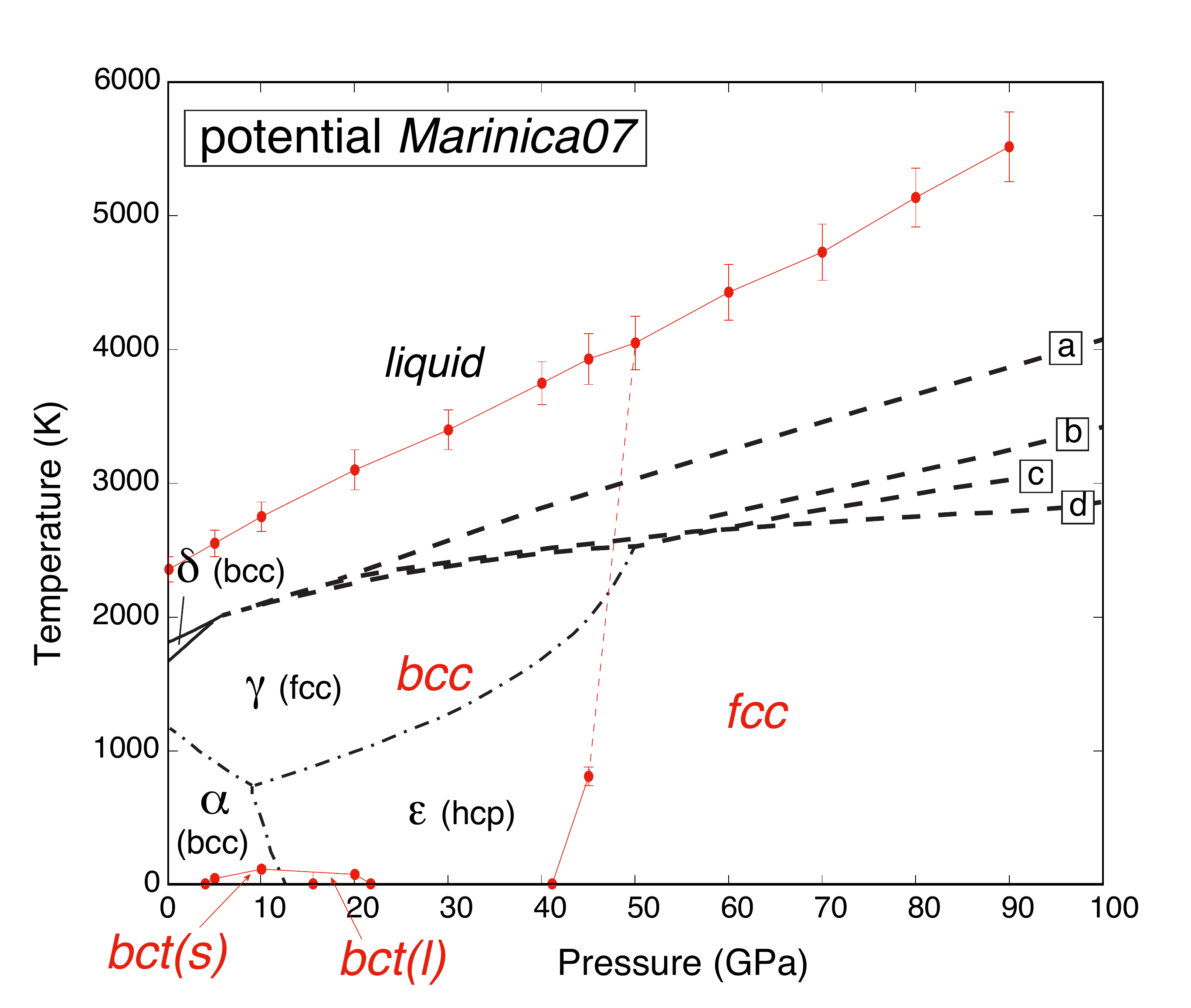}
\end{center}
\vspace{-20pt}
\caption {Phase diagram of iron with the Marinica07 potential. Black lines correspond to experimental data (long dash lines correspond to melting data: (a)\cite{Williams_melt_exp},  (b)\cite{static_melt_exp,static_melt_exp2,static_melt_exp3}, (c)\cite{Saxena_melt_exp} and (d)\cite{ge_triplepoint},while solid-solid transitions are shown by dash-dot lines~\cite{Yoo_solid_exp} and solid lines~\cite{Strong_delta_exp}. Suspected $\epsilon'$ and $\beta$ phases are not shown). Red symbols and lines correspond to the Marinica07 potential results.}
\label{fig:Malerba_pd}
\end{figure}

The phase diagram of the Marinica07 potential model is shown in Figure~\ref{fig:Malerba_pd}. 
The potential overestimates the melting temperature, at small pressure by 540~K, but the difference increases further with increasing pressure. 

The stable phase at lower pressures is bcc, with two different bct structures being the ground state below 100~K in the pressure range $4.0-21.7$~GPa.
Between 4~GPa and 15.5~GPa a shortened bct ($c/a=0.93$ with 0.37\% increase in density) is stable, while between 15.5~GPa and 21.7~GPa an elongated bct ($c/a=1.05$) is the ground state structure. 
Above 41.1~GPa the fcc structure becomes more stable than the bcc.


\subsection{potential Chamati06}

\begin{figure}[hbt]
\begin{center}
\includegraphics[width=7cm]{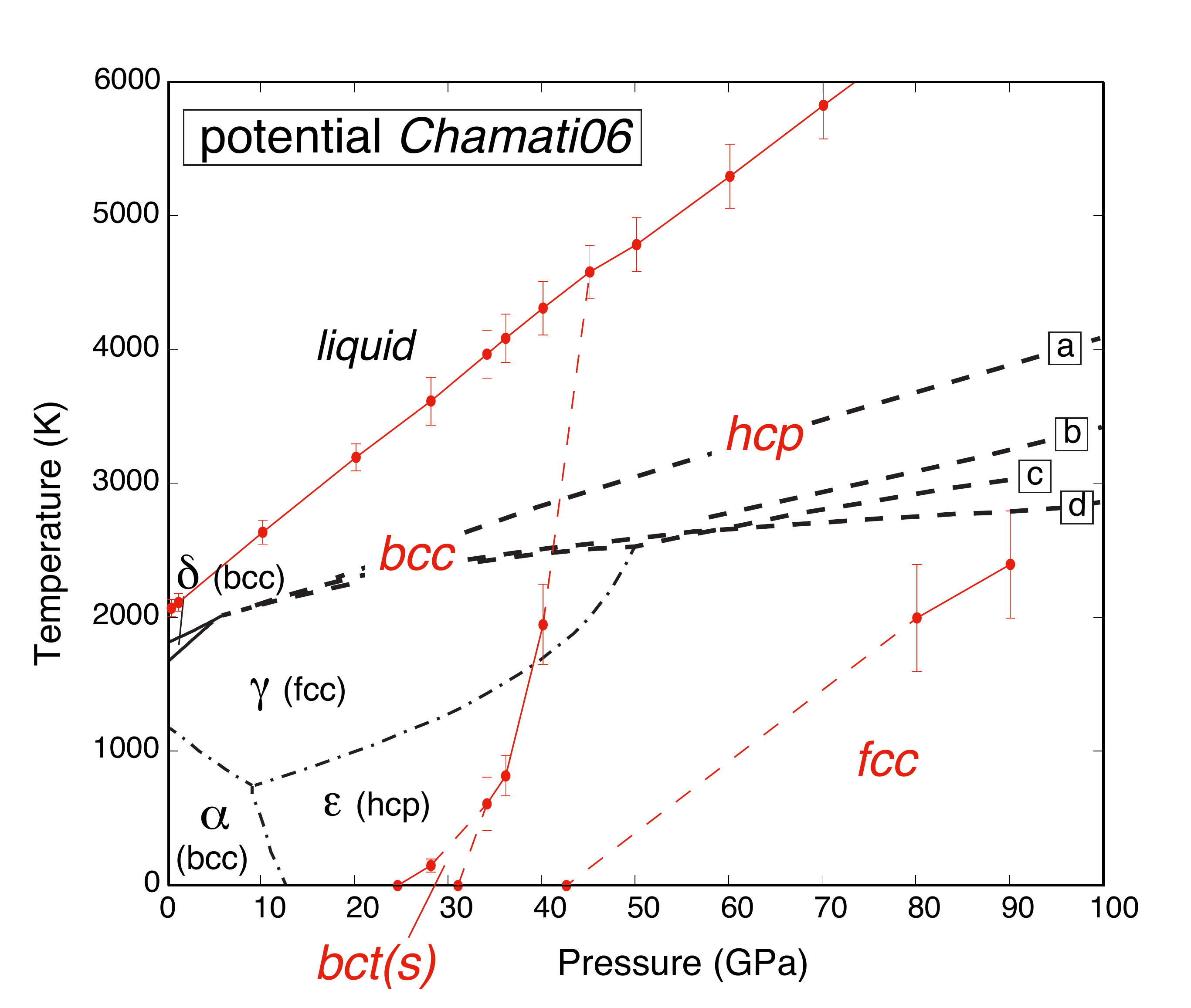}
\end{center}
\vspace{-20pt}
\caption {Phase diagram of iron with the Chamati06 potential. Black lines correspond to experimental data (long dash lines correspond to melting data: (a)\cite{Williams_melt_exp},  (b)\cite{static_melt_exp,static_melt_exp2,static_melt_exp3}, (c)\cite{Saxena_melt_exp} and (d)\cite{ge_triplepoint},while solid-solid transitions are shown by dash-dot lines~\cite{Yoo_solid_exp} and solid lines~\cite{Strong_delta_exp}. Suspected $\epsilon'$ and $\beta$ phases are not shown). Red symbols and lines correspond to the Chamati06 potential results.}
\label{fig:Chamati_pd}
\end{figure}

The phase diagram of the Chamati06 potential model is shown in Figure~\ref{fig:Chamati_pd}.
This potential overestimates the melting point by only $250$~K at low pressures, however the predicted melting line is very steep, the phase transition temperature becomes more than the double of the value suggested by experiments at 80~GPa.
As expected, the bcc is the low pressure stable phase, and there is again a narrow pressure range (between $25-31$~GPa) where the bct(s) phase ($c/a=0.92$ )becomes the most favourable at very low temperatures. 
Surprisingly, although the potential was fitted only to bcc and fcc parameters, the hcp structure becomes the stable structure at higher pressures.
Although the fcc becomes the ground state structure above 42~GPa, hcp continues to be the high temperature solid phase also for much higher pressures. 
Moreover, I found that a large number of walkers are needed to find and sample the fcc basin adequately, even at high pressures, suggesting that the phase space volume ratio of the fcc structure is very small compared to that of the hcp above transition.

\begin{figure}[hbt]
\begin{center}
\includegraphics[width=5.5cm,angle=90]{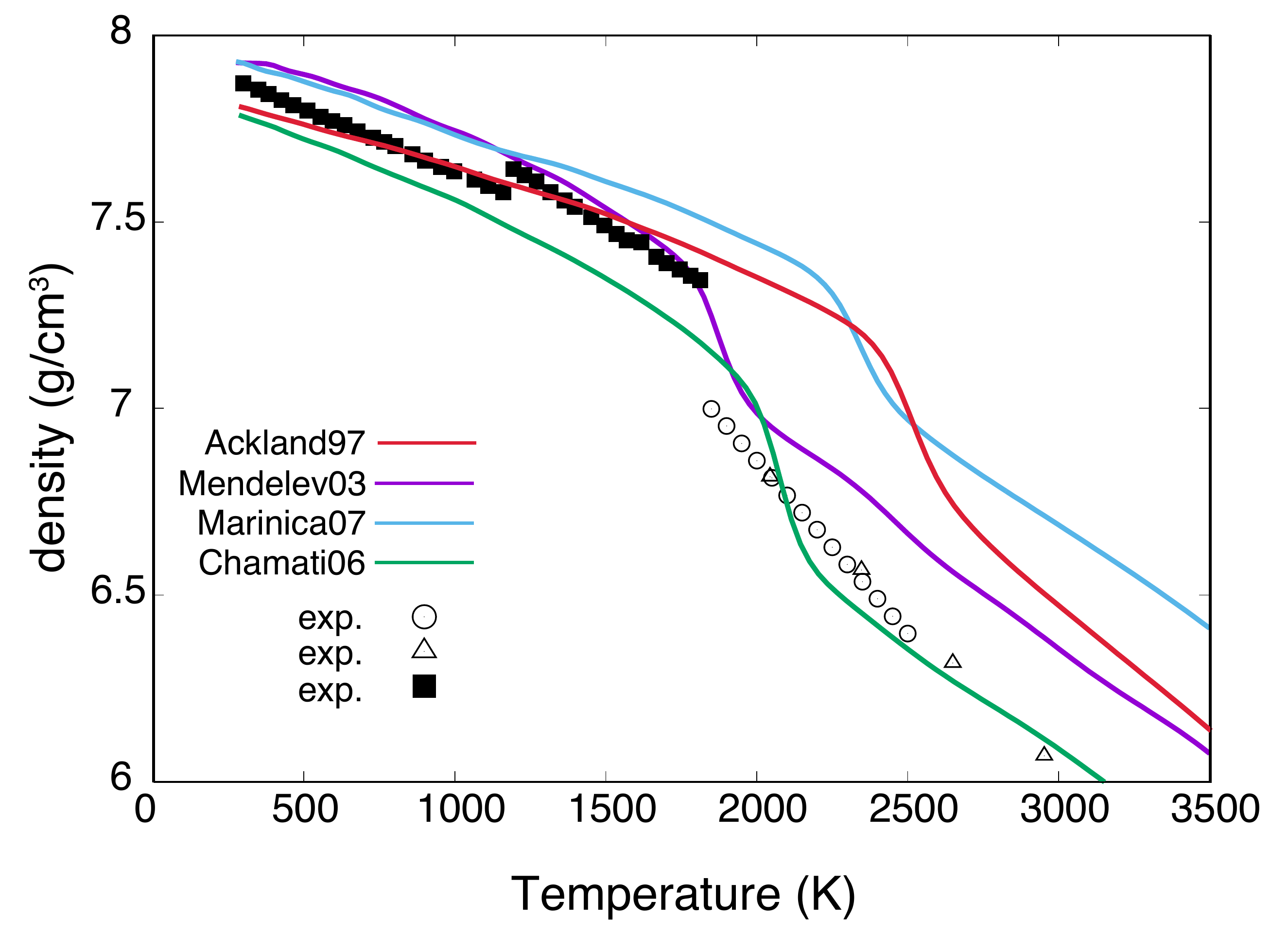}
\end{center}
\vspace{-20pt}
\caption {Density of iron as predicted by the studied models, as a function of temperature. Symbols correspond to experimental data, solid squares shows the density of solid iron (Ref.~\cite{solid_dens}), open circles and triangles correspond to liquid data (Ref.~\cite{liq_dens_data} and~\cite{liq_dens_data2}, respectively)}
\label{fig:density}
\end{figure}

Figure~\ref{fig:density} shows the density predicted by the four models at $p=0.1$~GPa as a function of temperature. 
Since the only stable structure of the models at low pressure is the bcc, we cannot expect them to show the density change corresponding to the $\alpha$-Fe to $\gamma$-Fe transition at 1183~K, which causes an expansion in the unit cell volume.
The density of the low temperature solid phase matches the experimental values relatively well, but the discrepancy slightly increases closer to the phase transition temperature. 
The density of the liquid phase shows a great variation and although its change with temperature shows a similar trend to that of the experimental line, the actual values are all larger, except for the Chamati06 potential.

\section{Conclusions}

The current work has demonstrated how the nested sampling method can be used to calculate the pressure-temperature phase diagram, without a prior knowledge of the phases, on the example of four interatomic potential models of iron.  
I have compared the predicted features of the phases to study the reliability of the models.
The bcc structure was found to be the low pressure stable phase for all four potentials, as one could presume from the fitting procedures of these models.
As the studied potentials do not describe the magnetic properties of iron, they cannot be expected to reproduce the low pressure phase behaviour where magnetism strongly influences the relative stability of polymorphs \cite{Bagno_Fe,magn_hcp}.
However, at higher pressures all the potentials have multiple stable phases, except the Ackland97 model.
The other three models have a pressure range where the body centred tetragonal phase is the most stable at very low temperatures, and the bcc phase transforms to either fcc or hcp at an even higher pressure. 
The melting temperature is considerably overestimated by three of the models, but the Mendelev03 potential shows a very good agreement with the melting line measured by static experiments~\cite{Williams_melt_exp}, not only at low pressure where the potential parameters were fitted, but at 90~GPa as well. 
This suggests that including low pressure liquid properties in the potential development process can significantly improve the melting behaviour of the potential also at high pressures, indicating that this improvement is transferable.

\section*{Acknowledgement}
The author thanks G\'abor Cs\'anyi for useful discussions, and Yuri Mishin and Lucas Hale for providing the potential files for their model, Chamati06.
LBP acknowledges support from the Royal Society through a Dorothy Hodgkin Research Fellowship.

\section*{Data Availability}
The raw data required to reproduce these findings are available to download from http://dx.doi.org/10.17632/kx832htkkv.1. The processed data required to reproduce these findings are available to download from http://dx.doi.org/10.17632/kx832htkkv.1.


\bibliographystyle{elsarticle-num.bst}
\bibliography{EAM_iron_references}

\begin{thebibliography}{10}
\expandafter\ifx\csname url\endcsname\relax
  \def\url#1{\texttt{#1}}\fi
\expandafter\ifx\csname urlprefix\endcsname\relax\def\urlprefix{URL }\fi
\expandafter\ifx\csname href\endcsname\relax
  \def\href#1#2{#2} \def\path#1{#1}\fi

\bibitem{bib:skilling}
J.~Skilling, Bayesian inference and maximum entropy methods in science and
  engineering, in: AIP Conference Proceedings, Vol. 735, 2004, p. 395.

\bibitem{bib:skilling2}
J.~Skilling, Nested sampling for general bayesian computation, J. of Bayesian
  Analysis 1 (2006) 833.

\bibitem{our_NS_paper}
L.~B. P\'artay, A.~P. Bart\'ok, G.~Cs\'anyi, Efficient sampling of atomic
  configurational spaces, J. Phys. Chem. B 114 (2010) 10502--10512.

\bibitem{our_NSHS_paper}
L.~B. P\'artay, A.~P. Bart\'ok, G.~Cs\'anyi, Nested sampling for materials: The
  case of hard spheres, Phys. Rev. E 89 (2014) 022302.

\bibitem{diffns}
B.~J. Brewer, L.~B. P\'artay, G.~Cs\'anyi, Diffusive nested sampling,
  Statistics and Computing 21 (2010) 649--656.

\bibitem{Frenkel_NS}
S.~Martiniani, J.~D. Stevenson, D.~J. Wales, D.~Frenkel, Superposition enhanced
  nested sampling, Phys. Rev. X 4 (2014) 031034.

\bibitem{pt_phase_dias_ns}
R.~J.~N. Baldock, L.~B. P\'artay, A.~P. Bart\'ok, M.~C. Payne, G.~Cs\'anyi,
  Determining pressure-temperature phase diagrams of materials, Phys. Rev. B 93
  (2016) 174108.

\bibitem{pymatnest_paper}
R.~Baldock, N.~Bernstein, K.~M. Salerno, L.~B. P\'artay, G.~Cs\'anyi, Constant
  pressure nested sampling with atomistic dynamics, Phys. Rev. E. 96 (2017)
  043311.

\bibitem{Strong_delta_exp}
H.~M. Strong, R.~E. Tuft, R.~E. Hanneman, The iron fusion curve and triple
  point, Metal. Trans. 4 (1973) 2657--2661.

\bibitem{Fe_structure}
J.~Donohue, Wiley, New York, 1974.

\bibitem{Yoo_solid_exp}
C.~S. Yoo, J.~Akella, A.~J. Campbell, H.~K. Mao, R.~J. Hemley, Phase diagram of
  iron by in situ x-ray diffraction: implications for earth's core, Science 270
  (1995) 1473.

\bibitem{ge_triplepoint}
R.~Boehler, in the earth's core from melting-point measurements of iron at high
  static pressures, Nature 363 (1993) 534.

\bibitem{dhcp_Fe2}
A.~S. Mikhaylushkin, S.~I. Simak, L.~Dubrovinsky, N.~Dubrovinskaia,
  B.~Johansson, I.~A. Abrikosov, Pure iron compressed and heated to extreme
  conditions, Phys. Rev. Lett. 99 (2007) 165505.

\bibitem{dhcp_Fe1}
C.~Yoo, P.~S\"{o}derlind, J.~Moriarty, A.~Cambell, dhcp as a possible new e’
  phase of iron at high pressures and temperatures of their electronic
  properties, Phys. Lett. A 214 (1996) 65--70.

\bibitem{Saxena_melt_exp}
S.~K. Saxena, G.~Shen, P.~Lazor, Experimental evidence for a new iron phase and
  implications for earth's core, Science 260 (1993) 1312.

\bibitem{Saxena_melt2_exp}
S.~K. Saxena, G.~Shen, P.~Lazor, Temperatures in earth's core based on melting
  and phase transformation experiments on iron, Science 264 (1994) 405.

\bibitem{Williams_melt_exp}
Q.~Williams, R.~Jeanloz, J.~Bass, B.~Svendsen, T.~J. Ahrens, The melting curve
  of iron to 250 gigapascals: A constraint on the temperature at earth's
  center, Science 236 (1987) 181--182.

\bibitem{static_melt_exp}
Y.-H. Sun, H.-J. Huang, F.-S. Liu, M.-X. Yang, F.-Q. Jing, A direct comparison
  between static and dynamic melting temperature determinations below 100 gpa,
  Chin. Phys. Lett. 8 (2005) 2002.

\bibitem{static_melt_exp2}
G.~Shen, H.~Mao, R.~J. Hemley, T.~S. Duffy, M.~L. Rivers, Melting and crystal
  structure of iron at high pressures and temperatures, Geophys. Res. Lett. 25
  (1998) 373.

\bibitem{static_melt_exp3}
Y.~Z. Ma, M.~Somayazulu, G.~Shen, H.~Mao, J.~Shu, R.~J. Hemley, In situ x-ray
  diffraction studies of iron to earth-core conditions, Phys. Earth Planet.
  Inter. 143 (2004) 455.

\bibitem{melt_Fe}
A.~Aitta, Iron melting curve with a tricritical point, J. Stat. Mech. 12 (2006)
  P12015.

\bibitem{ironpot_osetsky}
Y.~N. Osetsky, A.~G. Mikhin, A.~Serra, Study of copper precipitates in
  α‐iron by computer simulation i. interatomic potentials and properties of
  fe and cu, Phil. Mag. A 72 (1995) 361.

\bibitem{ironpot_Lee}
B.~J. Lee, M.~I. Baskes, H.~Kim, Y.~K. Cho, Second nearest-neighbor modified
  embedded atom method potentials for bcc transition metals, Phys. Rev. B 64
  (2001) 184102.

\bibitem{Ackland_potential}
G.~J. Ackland, D.~J. Bacon, A.~F. Calder, T.~Harry, Computer simulation of
  point defect properties in dilute fe—cu alloy using a many-body interatomic
  potential, Phil Mag A 75 (1997) 713--732.

\bibitem{MM_potential}
M.~I. Mendelev, S.~Han, D.~J. Srolovitz, G.~J. Ackland, D.~Y. Sun, M.~Asta,
  Development of new interatomic potentials appropriate for crystalline and
  liquid iron, Phil Mag A 83 (2003) 3977--3994.

\bibitem{Chamati_potential}
H.~Chamati, N.~Papanicolaou, Y.~Mishin, D.~A. Papaconstantopoulos,
  Embedded-atom potential for fe and its application to self-diffusion on
  fe(100), Surf. Sci. 600 (2006) 1793--1803.

\bibitem{Malerba_potential1}
L.~Malerba, M.-C. Marinica, N.~Anento, C.~Björkas, H.~Nguyen, C.~Domain,
  F.~Djurabekova, P.~Olsson, K.~Nordlund, A.~Serra, D.~Terentyev, F.~Willaime,
  C.~Becquart, Comparison of empirical interatomic potentials for iron applied
  to radiation damage studies, J. Nuclear Mat. 406 (2010) 19--38.

\bibitem{Malerba_potential2}
M.-C. Marinica, F.~Willaim, J.-P. Crocombette, Irradiation-induced formation of
  nanocrystallites with c15 laves phase structure in bcc iron, Phys. Rev. Lett.
  108 (2012) 025501.

\bibitem{Bagno_Fe}
P.~Bagno, O.~Jepsen, O.~Gunnarsson, Ground-state properties of third-row
  elements with nonlocal density functionals, Phys. Rev. B 40 (1997) R.

\bibitem{magn_hcp}
G.~Steinle-Neumann, L.~Stixrude, R.~E. Cohen, Magnetism in dense hexagonal
  iron, PNAS 101 (2004) 33--36.

\bibitem{ironpot_genrich}
G.~L. Krasko, B.~Rice, S.~Yip, A bond-order potential for atomistic simulations
  in iron, J. Computer-Aided Mat. Design 6 (1999) 129--136.

\bibitem{ironpot_muller}
M.~Müller, P.~Erhart, K.~Albe, Analytic bond-order potential for bcc and fcc
  iron—comparison with established embedded-atom method potentials, J. Phys.:
  Condens. Matter 19 (2007) 326220.

\bibitem{ironpot_mrovec}
M.~Mrovec, D.~Nguyen-Manh, C.~Els\"asser, P.~Gumbsch, Magnetic bond-order
  potential for iron, Phys. Rev. Lett. 106 (2011) 246402.

\bibitem{pymatnest}
N.~Bernstein, R.~J.~N. Baldock, L.~B. P\'artay, J.~R. Kermode, T.~D. Daff,
  A.~P. Bart\'ok, G.~Cs\'anyi, pymatnest,
  \url{https://github.com/libAtoms/pymatnest} (2016).

\bibitem{LAMMPS}
S.~Plimpton, Fast parallel algorithms for short-range molecular dynamics,
  Journal of computational physics 117~(1) (1995) 1--19.

\bibitem{bib:Q6parameter}
P.~J. Steinhardt, D.~R. Nelson, M.~Ronchetti, Bond-orientational order in
  liquids and glasses, Phys. Rev. B 28 (1983) 784.

\bibitem{iron_melt_interf}
D.~Y. Sun, M.~Asta, J.~J. Hoyt, Crystal-melt interfacial free energies and
  mobilities in fcc and bcc fe, Phys. Rev. B 69 (2004) 174103.

\bibitem{solid_dens}
F.~Tesfaye, P.~Taskinen, Densities of molten and solid alloys of (fe, cu, ni,
  co)-s at elevated temperatures-literature review and analysis, Tech. rep.,
  Dept. Mater. Sci. Eng., Aalto Univ., Espoo, Finland (2010).

\bibitem{liq_dens_data}
M.~J. Assaela, K.~Kakosimos, R.~M. Banish, J.~Brillo, I.~Egry, R.~Brooks, P.~N.
  Quested, K.~C. Mills, A.~Nagashima, Y.~Sato, W.~A. Wakeham, Reference data
  for the density and viscosity of liquid aluminum and liquid iron, J. Phys.
  Chem. Ref. Data 35 (2006) 285--300.

\bibitem{liq_dens_data2}
M.~Beutl, G.~Pottlacher, H.~J\"{a}ger, Thermophysical properties of liquid
  iron, Int. J. Thermophys. 15 (1994) 1323.

\end{thebibliography}

\end{document}